\documentclass[12pt,notitlepage,a4paper]{article}
\pdfoutput=1
\usepackage{color,graphicx}
\usepackage{cite}
\usepackage{amssymb}
\usepackage{delarray,amsmath}
\usepackage{cite}
\usepackage[table]{xcolor}
\usepackage[top=30mm, bottom=30mm, right=25mm, left=25mm, headheight=15.72pt]{geometry}

\title{\bf{Phonon and Shifton from a Real Modulated Scalar}}
\date{}
\author{Daniele Musso$^{a,}$\footnote{daniele.musso@usc.es}~ and Daniel Naegels$^{b,}$\footnote{daniel.naegels@ulb.ac.be}}

\begin{document}
\maketitle
\begin{center}\it{
$^{a}$Universidad de Santiago de Compostela (USC) and\\
Instituto Galego de F\'\i sica de Altas Enerx\'\i as (IGFAE)\\
\vspace{15pt}
$^{b}$Universit\'e Libre de Bruxelles (ULB) and\\
International Solvay Institute, Brussels.}
\end{center}
\vspace{30pt}

\begin{abstract}
We study a massive real scalar field that breaks translation symmetry dynamically.
Higher-gradient terms favour modulated configurations and neither finite density nor temperature are needed.
In the broken phase, the energy density depends on the spatial position and the linear fluctuations show phononic dispersion.
We then study a related massless scalar model where the modulated vacua break also the field shift symmetry and give rise to an additional
Nambu-Goldstone mode, the shifton. We discuss the independence of the shifton and the phonon and draw an analogy to rotons in superfluids.
Proceeding from first-principles, we re-obtain and generalise some standard results for one-dimensional lattices.
Eventually, we prove stability against geometric deformations extending existing analyses for elastic media 
to the higher-derivatives cases.

\end{abstract}
\newpage
\tableofcontents

\newpage

\section{Motivation and main results}

For about one hundred years now \cite{kojevnikov1993paul}, 
translation symmetry breaking has been extensively studied, 
though its dynamical origin is often left aside and neglected.
In typical condensed matter circumstances, 
there is a large hierarchy in energy between the physics of a crystal formation/melting and its low-energy excitations. 
These latter determine the thermodynamic and linear response properties,
which can be usually described by low-energy effective theories without considering the dynamical origin of the lattice.
Nevertheless, the physics of spatially-modulated order parameters,
as well as some conceptual questions concerning translation symmetry breaking,
require a dynamical and first-principle treatment.

The present paper aims at shedding light on some aspects of spontaneously broken translations relying on a simple field theory model.
We show that a real scalar field is enough to provide a setup where relevant questions can be directly addressed.
The main results are the following. 

\begin{enumerate}
 \item 
We prove by means of explicit examples that translations can be broken dynamically in field theory without the need of a finite density, 
a chemical potential or finite temperature.

\item
We consider models which enjoy a shift symmetry and cases where this is broken by a mass term and show that the spontaneous breaking of translations can 
be attained in both cases. Thereby, we show that shift symmetry is not a necessary ingredient to the purpose of describing phononic modes. 
More generically, we get spontaneous and inhomogeneous vacua where the energy density is spatially modulated and no auxiliary global symmetry is invoked\footnote{Unlike Q-lattice models, there is no internal global symmetry which leads to an homogeneous symmetry breaking of translation.}.

\item
In the shift-symmetric real scalar models that we analyse, 
the shift symmetry is spontaneously broken concomitantly with spatial translations.
We argue that at low-energy the Nambu-Goldstone modes associated to the two symmetries, \emph{e.g.} the \emph{shifton} and the \emph{phonon}, 
are well separated in momentum space and can therefore be regarded as effectively independent modes,
in analogy to what happens for (superfluid) phonons and rotons in superfluid helium 
\cite{Landini,2017PhRvL.119z0402C,Schmitt:2014eka,Nicolis:2017eqo}.
\end{enumerate}

\subsection{Context and method}

Recent effort in holographic models applied to condensed matter has focused on translation symmetry breaking \cite{Vegh:2013sk,Davison:2013jba,Blake:2013bqa,Davison:2013txa,Amoretti:2014zha,Amoretti:2014mma,Baggioli:2014roa,Alberte:2015isw}.
The possibility of striped phases \cite{Nakamura:2009tf,Donos:2011bh} triggered the quest for treatable modulated solutions, which often relies on 
the presence of an auxiliary global bulk symmetry \cite{Donos:2011ff,Andrade:2013gsa,Donos:2013eha}. In these cases, the spatial points of the broken phase 
are all equivalent modulo a global transformation and the translation breaking is \emph{homogeneous} \cite{Musso:2018wbv}, namely, the conserved densities do not 
depend on the spatial coordinates. Homogeneity entails valuable technical simplifications but it opens conceptual puzzles too. Homogeneous models have 
a trivial 1-point Ward-Takahashi identity for translations also when these are explicitly broken \cite{Amoretti:2016bxs}. To the purpose of realising and studying
cases where translations are broken inhomogeneously, we search for the simplest field theoretical model: we consider a real scalar field theory (introduced in Section 
\ref{MOD}) where the higher-gradient terms lead to a ``gradient Mexican hat'' that energetically favours spatially modulated solutions \cite{Musso:2018wbv}.
By means of a simple cosinusoidal ansatz \eqref{ansa}, 
we have analytical control on the harmonic modulated solutions without requiring to constrain the Lagrangian \eqref{model1}
(see Subsection \ref{tuning}).

The implementation of a gradient Mexican hat in the Lagrangian defines an enriched effective theory where translations are broken dynamically.
The non-trivial vacuum is explicitly obtained instead of being just assumed \emph{a priori}; similarly, the low-energy physics and the Nambu-Goldstone modes emerge from a dynamical
study without extra \emph{ad hoc} hypotheses. 
This allows one to realise the low-energy theorems on symmetry-breaking by construction
and to address the counting of Nambu-Goldstone modes in a direct fashion (Subsection \ref{tra_shi}).%
\footnote{On the problem of Nambu-Goldstone counting for spacetime symmetries we refer to \cite{Low:2001bw,Watanabe:2012hr,Kapustin:2012cr}}

In standard effective approaches, the role of shift symmetry is particularly important. 
First, standard effective field theories for solid, elastic media 
\cite{Leutwyler:1996er} and membranes rely on fields taking value in a physical target space;
translations symmetry in the target space corresponds to shift symmetry of the effective fields.
There, the spontaneous breaking of translations is actually a diagonal locking of translations and shifts leading to a single Nambu-Goldstone mode.
Second, shift symmetry allows one to study spacetime symmetry breaking systematically through coset constructions \cite{Nicolis:2013lma,Nicolis:2015sra}.%
\footnote{Third, the auxiliary global symmetry in the bulk of homogeneous holographic model can be thought in relation to shift symmetry of the effective low-energy theory,
see for example \cite{Amoretti:2017axe,Amoretti:2017frz,Amoretti:2018tzw,Amoretti:2019cef}.}
In the enriched effective approach we consider below, instead, shift symmetry is not necessarily present and hence it is not crucially related to the breaking of translations.
In fact, the shift symmetry we consider below is independent of translations and, when spontaneously broken, it leads to an additional Nambu-Goldstone mode which 
we call shifton.

The field $\phi$ in the Lagrangian \eqref{model1} does not represent the low-energy degrees of freedom emerging from the symmetry breaking
and thus $\phi$ is not directly identified with a Nambu-Goldstone. More precisely, the Nambu-Goldstone bosons are described by fluctuations 
of the field $\phi$ (as usual), however such fluctuations are considered around a non-trivial vacuum arising from a dynamical and spontaneous breaking.
As a result, the Lagrangian \eqref{model1} may (and will) contain terms where the scalar field appears without any derivative (as opposed to standard
effective theory constructions \cite{Son:2005rv}).

The present study belongs to a recently revived line of research about the treatment of translation symmetry breaking if field theory \cite{Watanabe:2011dk}, 
mainly in non-relativistic setups.%
\footnote{More broadly, the research field is related to the study of phase transitions by means of space-dependent configurations  \cite{Landau:1937obd,BOL1,BOL2}.}
In turns, translations are an aspect of a wider program concerning spontaneous and pseudo-spontaneous
symmetry breaking pursued in strongly coupled theories modelled holographically 
\cite{Argurio:2014rja,Argurio:2015via,Argurio:2015wgr,Argurio:2017irz,Argurio:2014uca,Bertolini:2015hua}.

\section{A real scalar model in $(1+1)$ dimensions}
\label{MOD}

We consider a canonical kinetic term, in particular we avoid higher time derivatives which would lead to Ostrogradsky instabilities \cite{Ostrogradsky:1850fid,Woodard:2015zca}.
We impose both spatial parity, $\partial_x \leftrightarrow -\partial_x$, and field-space parity, $\phi \leftrightarrow -\phi$. 
In an effective field theory spirit, we consider only terms up to the 4th order in $\phi$ and up to the 8th order in the spatial derivatives. Specifically, we take the model%
\footnote{The most general model fulfilling the requirements is given in Appendix \ref{gen_ter}.
The terms in \eqref{model1} provide a simple setup able to capture the translation symmetry breaking mechanism that constitutes the focus of the present analysis.}
\begin{align}\label{model1}
 {\cal L} =
 &\ \frac{1}{2} \dot\phi^2
 -\frac{A}{2} \phi'^2
 -\frac{m^2}{2} \phi^2
 +\frac{B}{4} \phi'^4
 +\frac{C}{2} \phi'^2 \phi''^2
 + D \phi''^4\ ,
\end{align}
where the dot indicates a time derivative while the prime denotes a derivative along the only spatial direction $x$.
We have considered a mass term so to break the rigid shifts $\phi \rightarrow \phi + c$ in the simplest possible way.
The equation of motion descending from \eqref{model1} is
\begin{equation}\label{ecu}
 \begin{split}
 \ddot\phi &= \phi'' \left(A-3 B \phi'^2+4 C \phi''' \phi' + 24 D \phi'''^2 \right)\\
 &\qquad +C \phi''^3 +C \phi'^2 \phi'''' +12 D \phi''^2\phi'''' -m^2 \phi \ .
 \end{split}
\end{equation}

We consider the static and harmonic ansatz
\begin{equation}\label{ansa}
 \phi(t,x) = \rho \cos(kx)\ ,
\end{equation}
characterised by a constant modulus $\rho$ and a constant wave-vector $k$.
Plugging the ansatz \eqref{ansa} into the equation of motion \eqref{ecu}, we get
\begin{equation}
 \begin{split}
 &3 k^4 \rho^2 \left[B+2k^2(C-6 D k^2)\right] \sin ^2(k x)\\ 
 & \qquad -\left(A k^2 +C k^6 \rho^2 - 12 D k^8 \rho^2 + m^2 \right)
 = 0\ ,
 \end{split}
\end{equation}
which is satisfied for
\begin{align}\label{eom_con_1}
 A &= -\frac{m^2}{k^2} - k^4 \rho^2 \left(C - 12 D k^2 \right) \ , \\ \label{eom_con_2}
 B &= -2k^2 (C-6 D k^2)\ .
\end{align}
Solving \eqref{eom_con_1} and \eqref{eom_con_2} in terms of $\rho$ and $k$, we obtain
\begin{align}\label{ro}
 \rho &= \pm \frac{\sqrt{2}}{B^{3/2}}\left[2(C^2+3B D)m^2 - A B C
 \right.\\ \nonumber & \left. \qquad \qquad
 \pm \frac{2 C (C^2+9BD)m^2 - AB(C^2+6B D)}{\sqrt{C^2+12 B D}}\right]^{1/2}\ ,\\ \label{ka}
 k &=\pm \left[\frac{C\pm\sqrt{C^2 + 12 B D}}{12 D}\right]^{1/2}\ .
\end{align}
We avoid a discussion of the radicands in \eqref{ro} and \eqref{ka} because in what follows we use \eqref{eom_con_1} and \eqref{eom_con_2} in the opposite direction,
and pick the theory with couplings $A$ and $B$ such that some chosen values of $\rho$ and $k$ provide a solution (we will in general take $\rho=k=1$).

Using the formul\ae$ $  of Appendix \ref{stress}, we compute the diagonal components of the stress-energy tensor for a solution \eqref{ansa},
\begin{equation}\label{ene_den}
 \begin{split}
 T_{tt} &= \frac{1}{8} \left\{4 \rho^2 \left(m^2 - 4 D k^8 \rho^2\right) \cos(2kx) \right. \\ 
 &\left. \qquad + k^6 \rho^4 \left[-C + 12 D k^2 + \left(C - 4D k^2 \right) \cos(4kx) \right] \right\}\ ,
 \end{split}
\end{equation}
and
\begin{equation}\label{Txx}
 T_{xx} = - \frac{1}{2} \rho^2 \left( m^2 + 6 D k^8 \rho^2 \right) \ .
\end{equation}

The model is invariant under translations, which translates into the Ward-Takahashi identity
\begin{equation}\label{1pt_WTi}
 \partial_\mu T^\mu_{\ \nu} = \partial_x T_{xx} = 0\ .
\end{equation}
Given the static character of the solution \eqref{ansa}, in order to satisfy the 1-point Ward-Takahashi \eqref{1pt_WTi}, the pressure $T_{xx}$ needs to be $x$-independent.
Nonetheless, the energy density $T_{tt}$ is spatially modulated. This constitutes an important generalisation with respect to ``Q-lattice" models where the energy density is spatially constant \cite{Musso:2018wbv}.

\subsection{Stability under geometric deformations}
\label{stage}

We generalise the analysis of static geometric deformations proposed in \cite{Domokos:2013xqa,Domokos:2013kha} to models with higher derivatives,
this provides a check%
\footnote{Notice that this check is necessary but not sufficient, stability is later confirmed relying on the analysis of the generic, time-dependent linear fluctuations.
The stability checks \cite{Domokos:2013xqa,Domokos:2013kha} are a generalisation of Derrick's theorem \cite{TopSol}.} 
of the stability of solution \eqref{ansa}.
We consider an infinitesimal geometric transformation parametrised by $\xi(x)$
\begin{equation}\label{geo_def}
 \bar x(x) = x + \xi(x)\ ;
\end{equation}
the bar will henceforth indicate transformed quantities in the sense of \eqref{geo_def}.
Since $\phi(x)$ is a scalar, it transforms as $\bar\phi(x) = \phi(\bar x)$.
The energy of the deformed system is given by%
\footnote{One can equivalently express it with respect to the deformed coordinates \eqref{geo_def},
\begin{equation}
 \begin{split}
 E[\bar \phi] &= \int dx\ {\cal E}\left[\bar \phi(x), \partial \bar\phi(x), \partial^2 \bar\phi(x)\right]\\
              &= \int d\bar x\ \left|\frac{\partial x}{\partial \bar x} \right|\ {\cal E}\left[ \phi(\bar x), \frac{\partial \bar x}{\partial x} \bar \partial \phi(\bar x), 
              \frac{\partial \bar x}{\partial x} \bar \partial\left(\frac{\partial \bar x}{\partial x} \bar \partial \phi(\bar x)\right)\right]\ .
 \end{split}
\end{equation}
}
\begin{equation}
 \begin{split}
 E[\bar \phi] &= \int dx\ {\cal E}\left[\bar \phi(x), \partial \bar\phi(x), \partial^2 \bar\phi(x)\right]\\
              &= \int dx\ {\cal E}\left[ \phi(\bar x(x)), \partial \phi(\bar x(x)), \partial^2 \phi(\bar x(x))\right]\ ,
 \end{split}
\end{equation}
and can be expanded in powers of $\xi$ (and its derivatives),
\begin{equation}
 E[\bar\phi] = E[\phi] + E_1[\phi] + E_2[\phi] + ...\ .
\end{equation}
The linear term $E_1[\phi]$ vanishes on-shell in the static limit.
By means of integrations by parts%
\footnote{We consider geometric deformations such that $\xi(x)$ has a compact support.}, 
the quadratic term $E_2 = \int  dx\ {\cal E}_2$ can be written in the ``diagonal'' form
\begin{equation}\label{E2}
 {\cal E}_2 = d_{22}\, (\partial^2\xi)^2 + d_{11}\, (\partial\xi)^2\ .
\end{equation}
The coefficients in \eqref{E2} are given by
\begin{align}\label{d22}
 d_{22}=&-\frac{1}{2} k^4 \rho ^4 \sin ^2(k x) \left[C \sin ^2(k x)+12 D k^2 \cos ^2(kx)\right]\ ,\\ \label{d11}
 d_{11}=&-\frac{1}{16} \rho ^2 \left[3 k^6 \rho ^2 \left(C-12 D k^2\right) \cos (4 k x)+k^6 \rho ^2 \left(C+228 D k^2\right) 
 \right. \\ & \left.
 -4 \cos (2 k x) \left(C k^6 \rho^2+m^2\right)+4 m^2\right]\ .
\end{align}
In order for the system to be stable against generic geometric deformations \eqref{geo_def}, we must require
\begin{equation}\label{test}
 d_{22}\geq 0 \ , \qquad
 d_{11}\geq 0 \ , 
\end{equation}
to hold locally (see Figure \ref{geo_stab} for an explicit example).

\subsection{Fluctuations}
\label{flu}

An infinitesimal spatial translation of the solution \eqref{ansa} is encoded in the following field variation
\begin{equation}\label{tra_zer_mod}
 \Delta_\xi \phi(x) = \xi \phi'(x) = \xi \rho k \sin(kx)\ ,
\end{equation}
in fact%
\footnote{
An operatorial rephrasing of \eqref{tra_zer_mod} is given by
\begin{equation}
\Delta \phi(x) = [P_x, \phi(x)]
= \frac{i}{\hbar}\int dy [\pi(y)\phi'(y),\phi(x)]
= \phi'(x)\ ,
\end{equation}
where $\pi(x)$ is the variable conjugated to $\phi(x)$
\begin{equation}\label{coni}
 [\pi(x),\phi(y)] = -i \hbar\, \delta(x-y)\ .
\end{equation}} 
\begin{equation}
\begin{split}
 &\phi(t,x-\xi) = \rho \cos[k(x-\xi)] \\ &
 \qquad \qquad = \rho \cos(kx) + \xi \rho k \sin(kx) +... = \phi(t,x) + \Delta_\xi \phi(x) + ...\ .
 \end{split}
\end{equation}
Translations connect degenerate solutions and \eqref{tra_zer_mod} is an infinitesimal 
motion along the associated zero mode given by $\phi'(x)$.
Let us take a brief but useful digression. One could define the fluctuations
\begin{equation}\label{digre}
\begin{split}
 \phi(t,x) &= \rho \cos(kx) + \xi(t,x)\, \phi'(x) \\
               &= \rho \cos(kx) + \xi(t,x)\, \rho k \sin(kx)\ ,
\end{split}
\end{equation}
promoting the parameter $\xi$ in \eqref{tra_zer_mod} to be a \emph{normalisable} function of spacetime;
the fluctuation field $\xi(t,x)$ would represent a modulation of the zero mode $\phi'(x)$.%
\footnote{This way of thinking is connected to more standard cases.
For instance, a flat D-brane breaks perpendicular translations spontaneously.
The corresponding zero mode is just an orthogonal rigid shift of the entire D-brane. 
The Nambu-Goldstone mode can be thought of as a normalisable modulation of such rigid shift \cite{Low:2001bw}. 
In our case, a rigid mode should be seen as a fluctuation of the form \eqref{tra_zer_mod} where $\xi$ is constant.
Note also that formula \eqref{digre} is analogous to the fluctuation parametrisation considered in \cite{Musso:2018wbv}
for a modulated complex scalar field background $e^{i k x}$,
\begin{equation}
 \delta \phi(t,x) = e^{i k x} \varphi(t,x)\ .
\end{equation}
Despite the formal similarity, this is not a Bloch wave, as the function 
$\varphi(t,x)$ does not in general have the spatial periodicity of the background.}
It is important to note that the translation zero-mode \eqref{tra_zer_mod} is characterised 
by the same wavevector $k$ of the background solution. Thus, we expect a massless mode at momentum
$k$ in Fourier space. This observation will prove useful below.
It is natural to compare the configuration \eqref{ansa} to a discrete chain of wavevector $k$,
namely ``sampling" the modulated continuous profile \eqref{ansa} at points separated by $\frac{2\pi}{k}$.
The deformation \eqref{tra_zer_mod} corresponds to an equal shift for all the sampling points 
and the discrete chain cannot distinguish between \eqref{tra_zer_mod} or a spatially constant deformation, 
the latter leading to massless phonons in a one-dimensional discrete lattice. 

To keep the calculations easier, instead of adopting \eqref{digre} we define the fluctuations
\begin{equation}
\begin{split}
 \phi(t,x) &= \rho \cos(kx) + \varphi(t,x)\ .
\end{split}
\end{equation}
The Lagrangian at linear order in the fluctuations is a total derivative, ${\cal L}_1 = {\cal B}'$, where
\begin{equation}
\begin{split}
 {\cal B} = -k^4 \rho^3 \cos(kx) \left(4D k^2 \cos(kx)^2 + C \sin(kx)^2\right) \varphi' 
 -\frac{m^2 \rho  \sin (k x) \varphi}{k}\ .
\end{split}
\end{equation}
The quadratic Lagrangian density is 
\begin{align}\label{qua}
{\cal L}_2 =&
\frac{1}{2} \dot\varphi^2
-\frac{1}{2} m^2 \varphi^2\\ \nonumber &
+\frac{2 m^2 - 3 k^6 \rho^2 (C-4 Dk^2)+ k^6 \rho^2 \cos(2kx)(7C-36Dk^2)}{4k^2}\varphi'^2
\\ \nonumber &
+\frac{1}{2} k^2 \rho^2 \left[C \sin^2(k x) +12 D k^2 \cos^2(kx)\right] \varphi''^2
+C k^3 \rho ^2 \sin (2 k x) \varphi' \varphi''\ .
\end{align}

The coefficients of the quadratic Lagrangian \eqref{qua} are space-dependent.
We can go to Fourier space but the various harmonic component of the fluctuation field are thereby mixed,
\begin{equation}\label{fur}
 \begin{split}
\tilde {\cal L}_2 =&
a_0(k,\omega,q) \tilde\varphi(-\omega,-q) \tilde\varphi(\omega,q)
+a_+(k,\omega,q) \tilde\varphi(-\omega,-2k-q) \tilde\varphi(\omega,q) \\ &
+a_-(k,\omega,q) \tilde\varphi(-\omega,2k-q) \tilde\varphi(\omega,q)\ .
 \end{split}
\end{equation}
The mixing occurs only among modes with fluctuation momenta that differs by $2k$.
In order to express the quadratic Fourier Lagrangian \eqref{fur} in a matrix form, we define the coefficient functions
\begin{align}\label{coe_fun}
a_0(k,\omega,q) &= \pi m^2 \left(\frac{q^2}{k^2}-1\right) \\ \nonumber
&+ \frac{\pi}{2} \left[C k^2 \rho^2 q^2  \left(q^2-3 k^2\right)
+12 D k^4 \rho^2 q^2 (k^2 + q^2)+2\omega^2\right]\ ,\\
a_+(k,\omega,q) &=\frac{\pi}{4} (C-12Dk^2)\, k^2 \rho^2\, q  (k-q) (2 k+q) (3 k+q)\ ,\\
a_-(k,\omega,q) &=-\frac{\pi}{4} (C-12Dk^2)\, k^2 \rho^2\, q  (2 k-q) (3 k-q) (k+q)\ ,
\end{align}
which satisfy the following symmetry properties
\begin{align}
 a_-(k,\omega,q) &= a_+(-k,\omega,q) \ ,\\ \label{hermite}
 a_+(k,\omega,q) &= a_-(k,\omega,q+2k)\ .
\end{align}
To avoid clutter, we henceforth simplify the notation $a(k,\omega,q)\to a(q)$ understanding the explicit dependence on $k$ and on $\omega$.
The infinite dimensional matrix associated to the quadratic form \eqref{fur} is 
\begin{equation}\label{matriciana}
  M= \left(\begin{array}{ccccccc}
... & ... & ... & ... & ... & ...& ...\\
... & a_0(q+4k) & a_+(q+2k) & 0 & 0 & 0 & ...\\
... & a_-(q+4k) & a_0(q+2k) & a_+(q) & 0 & 0 &...\\
... & 0 & a_-(q+2k) & a_0(q) & a_+(q-2k) & 0 & ...\\ 
... & 0 & 0 & a_-(q) & a_0(q-2k) & a_+(q-4k) & ...\\
... & 0 & 0 & 0 & a_-(q-2k) & a_0(q-4k) & ... \\
... & ... & ... & ... & ... & ...& ...
 \end{array} \right)
\end{equation}
Note that the property \eqref{hermite} implies $M^\dagger=M$.%
\footnote{Since at any row the momenta are shifted by different multiples of  $2k$, the matrix $M$ is not circulant.}

In order to deal with the infinite dimensional matrix \eqref{matriciana}, 
we actually consider $(2N+1)\times(2N+1)$ finite submatrices denoted as $M_{[N]}$
whose diagonal entries go from $a_0(q-2Nk)$ to $a_0(q+2Nk)$.%
\footnote{The matrix explicitly written in \eqref{matriciana} (dots excluded) would correspond to $M_{[2]}$.}
The possibility of taking finite-$N$ truncations of the matrix $M$ is later validated by the fact that the relevant results convergence quickly with respect to $N$,
becoming therefore soon insensitive to the truncation itself.

According to the observations made below Equation \eqref{digre}, 
we expect a massless mode about $q=k$, we then define $q=k+p$ where $p$ is small.%
\footnote{Since we want to capture the physics at $q\sim k$, we cannot rely on a mean-field approximated treatment
based on spatial averaging.}
 At $p=0$, $M_{[N]}$ has a $2\times 2$ $N$-independent diagonal block%
\begin{equation}\label{mino}
 \begin{split}
M_b =  &\left(
 \begin{array}{cc}
  a_0(\omega,k) & a_+(\omega,-k)\\
  a_-(\omega,k)& a_0(\omega,-k)
 \end{array}
 \right)
  =
 \pi\left(
 \begin{array}{cc}
  \omega^2 - \frac{m_1^2}{2} & -\frac{m_1^2}{2}\\
   -\frac{m_1^2}{2} & \omega^2 - \frac{m_1^2}{2} 
 \end{array}
 \right)\ ,
 \end{split}
\end{equation}
where
\begin{equation}\label{m1}
 m_1^2 = 2 k^6 \rho^2 (C-12 D k^2)\ .
\end{equation} 
From the determinant of the submatrix $M_b$ we get the masses of two modes
\begin{equation}
 \text{det}(M_b) = \pi^2 \omega^2 \left(\omega^2 - m_1^2\right)\ ,
 \label{detMb}
\end{equation}
namely $0$ and $m_1$. 
The massless mode can be interpreted as an (acoustic) phonon representing the Nambu-Goldstone mode of spontaneously broken translations. 
The gapped mode corresponds instead to an optical branch (its interpretation is commented in Subsection \ref{pc}).

Recalling \eqref{detMb}, the determinant of $M_{[N]}$ can be written as
\begin{equation}\label{dette}
 \begin{split}
& ({\cal A}\omega^2 + {\cal B}_{[N]} p^2 + {\cal C}_{[N]} p^4 + ...) \cdot 
({\cal D} + {\cal E} \omega^2 + {\cal F}_{[N]} p^2 + {\cal G}_{[N]} p^4 + ...) \\ & \qquad
 \cdot ({\cal H}_{[N]} + {\cal I}_{[N]} \omega^2 + {\cal L}_{[N]} p^2 + {\cal M}_{[N]} p^4 + ...) \cdot ...\ ,
 \end{split}
\end{equation}
where the coefficients ${\cal A}$, ${\cal D}$ and ${\cal E}$ do not depend on $N$, while the other coefficients do.
The squared propagation velocity of the acoustic phonon is given by
\begin{equation}
 \omega^2 = -\frac{{\cal B}_{[N]}}{{\cal A}} p^2 = c^2_{[N]} p^2 + ...\ ,
\end{equation}
and can be computed in terms of
\begin{equation}\label{speed}
c^2_{[N]} = -\frac{c^{(0,2)}_{[N]}}{c^{(2,0)}_{[N]}}\ ,
\end{equation}
where $c^{(n,m)}_{[N]}$ is defined to be the coefficient of the term $O(\omega^n,p^m)$ of the determinant \eqref{dette} of $M_{[N]}$.%
\footnote{One can compute also the $q^2$ term in the phonon dispersion relation.
From \eqref{dette} we have 
\begin{equation}
 \frac{{\cal C}_{[N]}}{{\cal A}} = \frac{c^{(0,4)}_{[N]}}{c^{(2,0)}_{[N]}} - \frac{c^{(0,2)}_{[N]}}{\left(c^{(2,0)}_{[N]}\right)^2} \left(c^{(2,2)}_{[N]} 
 - \frac{c^{(0,2)}_{[N]}}{c^{(2,0)}_{[N]}}c^{(4,0)}_{[N]}\right)\ .
\end{equation}}
Recall that in order to obtain the true phonon speed we would need to take the $N\rightarrow \infty$ limit of \eqref{speed}.
We show below that there exist cases where \eqref{speed} converges quickly (increasing the truncation level $N$) to a stable value.

\subsection{Explicit example}

\begin{figure}[t]
\begin{center}
\includegraphics[width=0.49\textwidth]{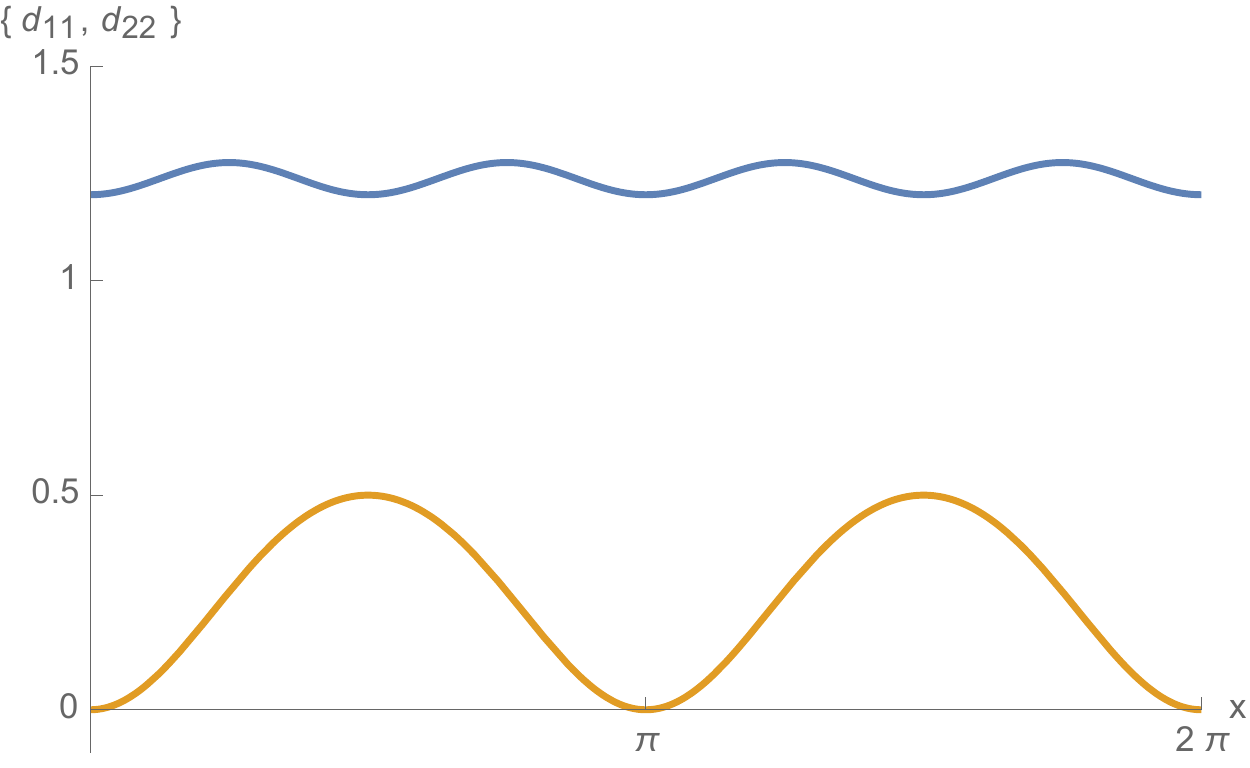}
\includegraphics[width=0.49\textwidth]{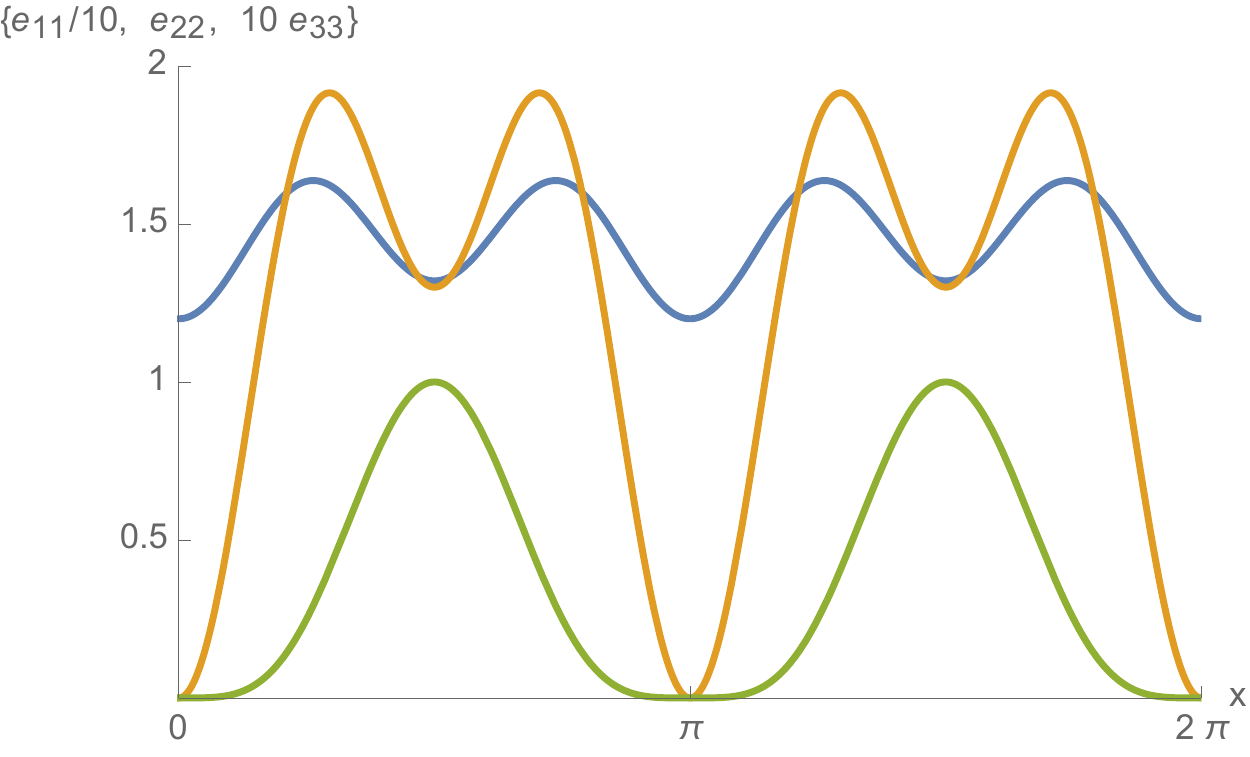}
\end{center}
\caption{Stability test with respect to static geometric deformations.
Left: Model \eqref{model1} in the particular case \eqref{cas_mas}, \eqref{karo};
the functions $d_{11}$ and $d_{22}$ are given in \eqref{d11} and \eqref{d22}.
Right: Model \eqref{model2} in the specific case \eqref{expl_ext1} and \eqref{expl_ext2};
the functions $e_{11}$, $e_{22}$ and $e_{33}$ are given in \eqref{d33_ext}.
All the plotted functions are positive across the entire unit cell, $x \in \{0, \frac{2\pi}{k}\}$ with $k=1$, so the specific cases considered pass the stability test.}
\label{geo_stab}
\end{figure}

Depending on the specific values of the coefficients in the Lagrangian \eqref{model1}, one can find either stable or unstable modulated solutions.
We consider a specific stable case, which is representative of a stability region within the coupling space, namely we take
\begin{equation}\label{cas_mas}
 m=1\ , \qquad C=-1\ , \qquad D = -\frac{1}{10}\ ,
\end{equation}
We fix the coefficients $A$ and $B$ in the Lagrangian \eqref{model1} according to \eqref{eom_con_1} and \eqref{eom_con_2} and requiring that the model admits a solution \eqref{ansa} with
\begin{equation}\label{karo}
 k=1\ , \qquad \rho = 1\ .
\end{equation}
These specific choices pass the stability check \eqref{test}, see Figure \ref{geo_stab}.

 According to the formula \eqref{speed}, we compute the phonon propagation speed $c_{[N]}$ for the first truncation levels $N=1,2,3,4,...$ 
 and see that it converges to a finite value in an extremely rapid fashion,%
 \footnote{In Appendix \ref{alte} we also describe an alternative approximated computation based on the recursion structure of $M$ (this latter described in Appendix \ref{rec_stru}).}
\begin{align}\label{converga}
 c_{[N]} &= \left\{\frac{21 \sqrt{\frac{67}{73}}}{10},\frac{3
   \sqrt{\frac{15326191}{2131235}}}{4},\frac{3
   \sqrt{\frac{96258350059}{2141686810}}}{10},\frac{9
   \sqrt{\frac{592084723513469}{1185615627221}}}{100},...\right\}\\ \nonumber
 &\sim \left\{2.01185,\ 2.01123,\ 2.01123,\ 2.01123,\ ...\right\}\ .
\end{align}
One should not worry about $c_{[N]}>1$,
in fact the model is non-relativistic from the start and the coefficient in front of the $\dot \phi^2$ in the Lagrangian \eqref{model1}
has been set to its canonical value without any precise physical argument; these could come from the study of relativistic UV completions.

\begin{figure}[t]
\begin{center}
\includegraphics[width=0.49\textwidth]{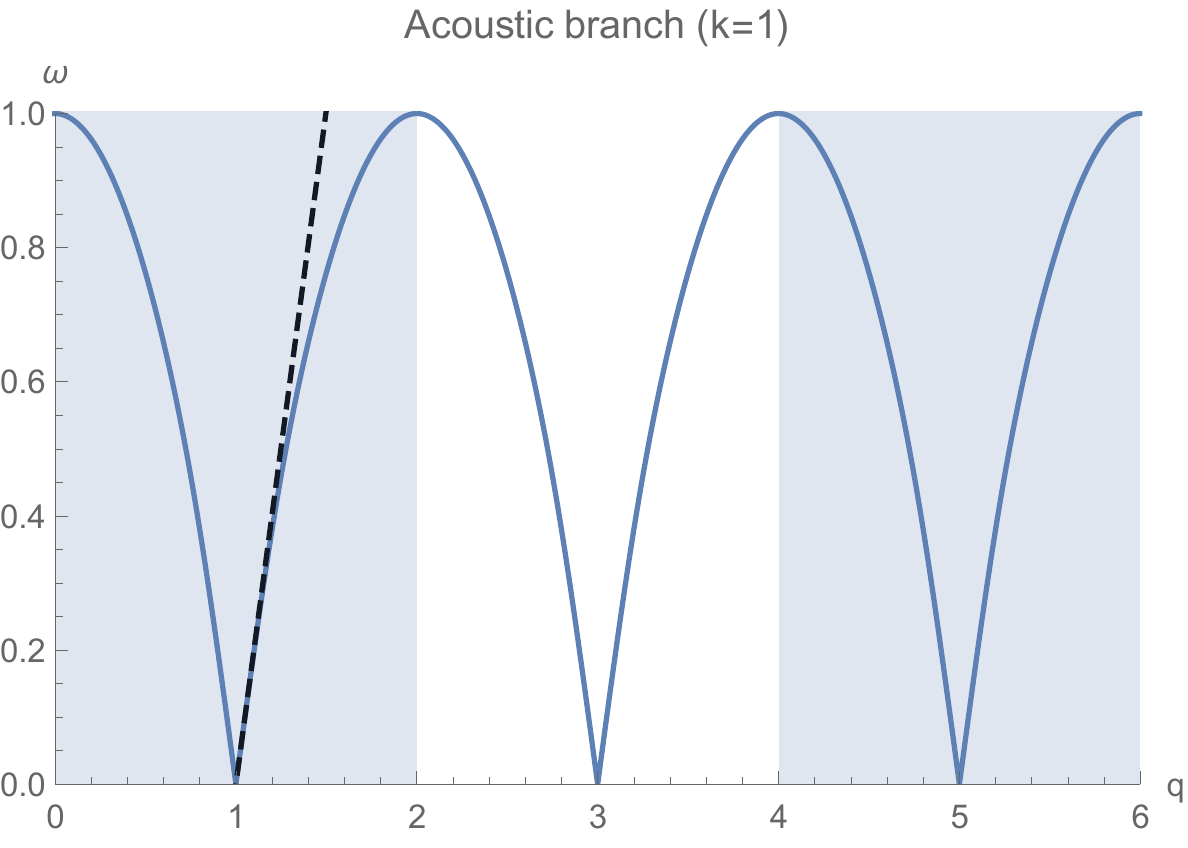}
\includegraphics[width=0.49\textwidth]{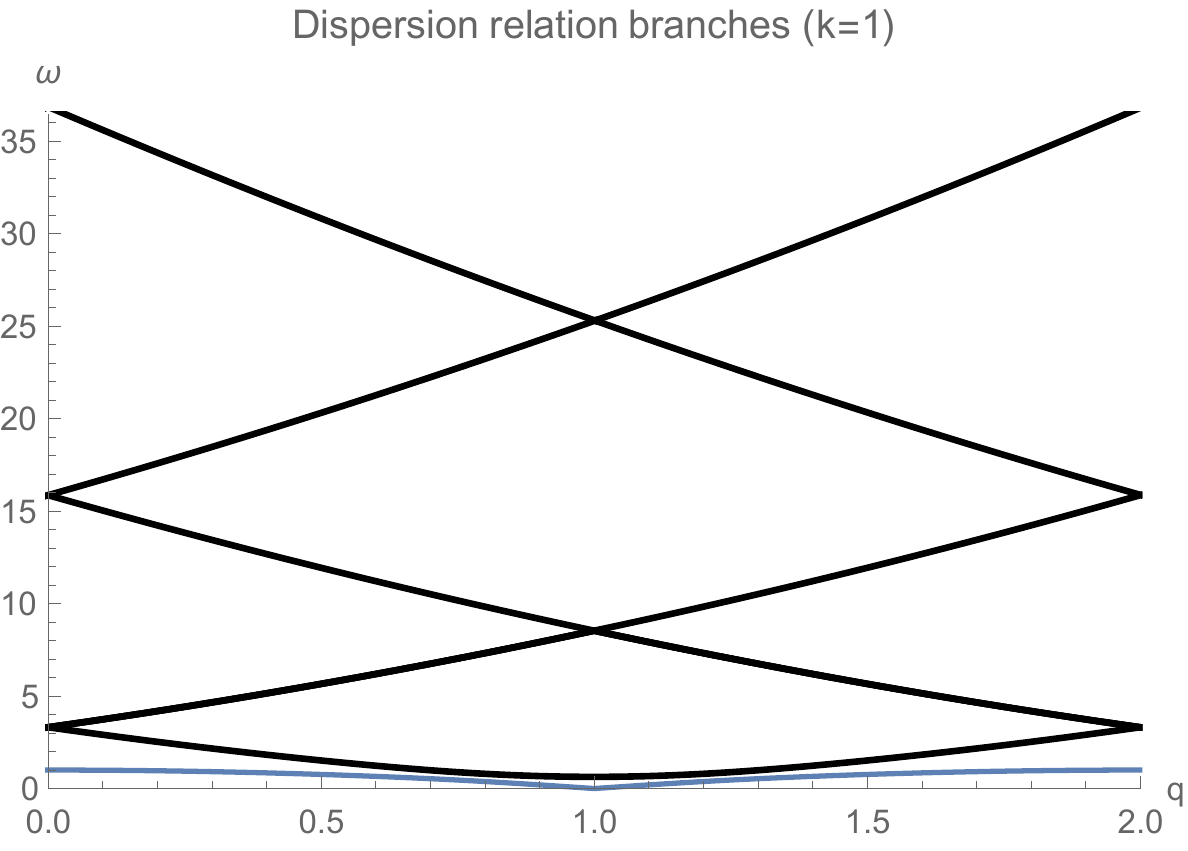}
\end{center}
\caption{Spectrum of the the linear fluctuations of model \eqref{model1} around a solution \eqref{ansa} (the plots refers to the specific case \eqref{cas_mas} and \eqref{karo}).
Left: lower bouncing branch corresponding to acoustic phonons; the dashed line indicates the phonon propagation speed for $q\sim k=1$.
Right: The blue line is again the same acoustic phonon branch of the left panel, the black lines are instead the optical branches; 
these latter have polynomial concave shape (see Figure \ref{com_cha}, right panel) 
and there is a branch for any $2k$ multiple.  }
\label{pho_dis}
\end{figure}

The study of the determinant of the matrix \eqref{matriciana} in the specific case \eqref{cas_mas} and \eqref{karo},
and in particular the (numerical) identification of the loci where it vanishes, allows us to find the complete 
linear fluctuation spectrum and have a full characterisation of the dispersion relations.
The whole mode structure features two kinds of dispersion relations: an acoustic ``bouncing" lower branch
and optical upper branches which are concave and repeated for any multiple of $2k$, see Figure \ref{pho_dis}.

The acoustic branch is analogous to the dispersion relation of the phonons of the classical one-dimensional chain (see Subsection \ref{cha});
note that the acoustic phonon branch features vanishing frequency at $q=k+2nk$ with $n$ a relative integer (and not at $q=0$).
At $q\sim k$ the dispersion relation is linear and corresponds to the propagation speed already evaluated analytically in \eqref{converga}.

The gap of the first optical branch is given by \eqref{m1} which, evaluated in the case \eqref{cas_mas} and \eqref{karo}, returns $m_1= 0.632...$ 
in agreement with the value found numerically.
The right panel of Figure \ref{pho_dis} shows the first Brillouin zone, which we define as $q\in [0,2k)$.

\subsection{Comparing to the one-dimensional chain}
\label{cha}

The eigenfrequencies of the discrete chain of balls and springs are given by (see for instance \cite{Ziman})
\begin{equation}\label{spr_bal}
 \omega(q) = 2 \sqrt{\frac{g}{\bar m}} \left|\sin\left(\frac{q}{2}\right)\right|\ ,
\end{equation}
where $\bar m$ is the mass of the balls and $g$ is the second derivative of the potential between two neighbouring balls.
The dispersion relation in \eqref{spr_bal} is taken from the chain with only one kind of atoms,
while we said above that our continuous model corresponds intuitively to a chain whose unit cell possess an infinite number of internal degrees of freedom. 
Recall, however, that the lowest phononic branch of a chain with any number of different atoms in the unit cell can be thought of as the mode of a chain with only one kind of atom;
in fact, the lowest mode corresponds to the unit cell moving rigidly without deforming. 
\begin{figure}[h]
\begin{center}
\includegraphics[width=0.49\textwidth]{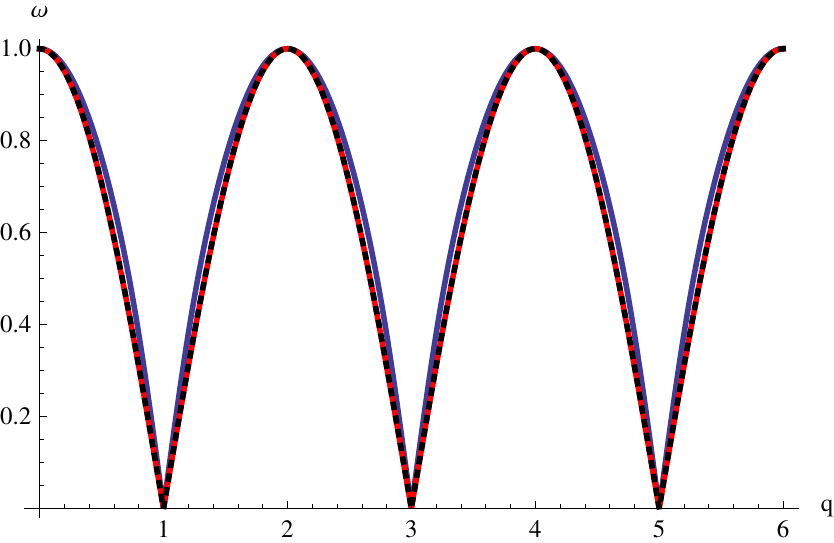}
\includegraphics[width=0.49\textwidth]{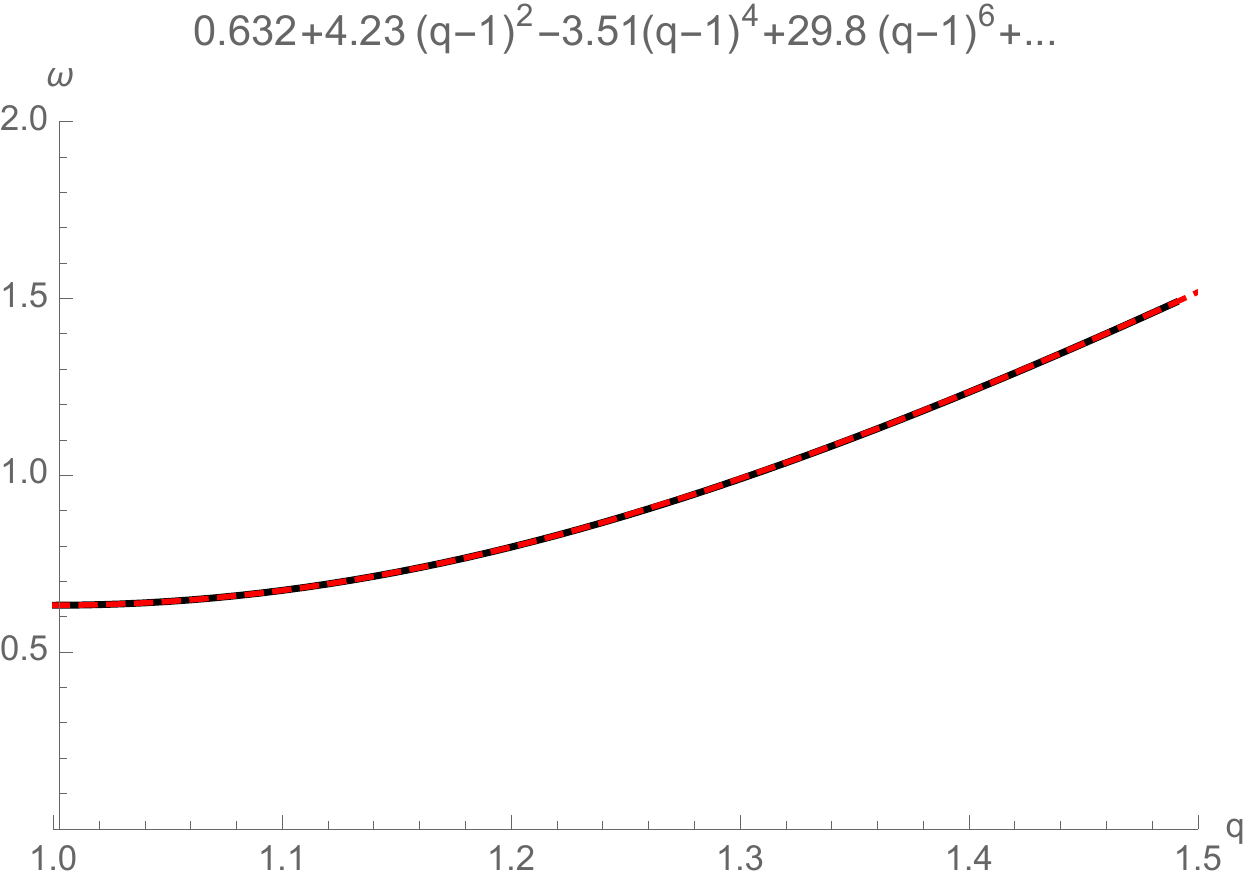}
\end{center}
\caption{Left: Comparison of the phonon dispersion relation of model \eqref{model1} with that of the one-dimensional chain:
the black dashed line corresponds to \eqref{spr_bal} with $q\rightarrow \pi(q-1)$ and $\frac{g}{\bar m}=\frac{1}{4}$;
the blue line is the phonon dispersion relation of Figure \ref{pho_dis};
the red line is the phonon dispersion relation obtained with $k=\rho=m=1$ and $C=-\frac{173}{50}$, $D=-2$, which approximates \eqref{spr_bal} to the $.001$ level.
Right: Even polynomial fit of the first optical branch centred at $q=1$.
The discrepancy between the numerical points and the fit is less than $10^{-10}$ in the whole figure range; the order of the fitting polynomial (red dashed) is $20$; 
the lowest coefficients have been observed to stabilise since polynomial order $16$.}
\label{com_cha}
\end{figure}
To emphasise this analogy with the discrete mono-atomic chain, we compare the two dispersion relations in Figure \ref{com_cha}. 
To push further the comparison to discrete chains, if we look to a chain with several types of atoms, the internal oscillations within a unit cell correspond to optical modes.
This might give a physical picture of the reason why we have a tower of optical modes in our continuous model.

\subsection{Comparing to a one-dimensional kink crystal}

An exactly tractable model of a one-dimensional solid can be built from a lattice array of kinks of a scalar field obeying the sine Gordon equation \cite{PhysRevA.8.2514}.
The Hamiltonian density is 
\begin{equation}\label{Hsg}
 {\cal H} = \frac{1}{2} \dot\phi^2 + \phi'^2 + 4\mu^2 \sin^2\left(\frac{1}{2}\phi\right)\ .
\end{equation}
The static solutions obey the equation of motion of a pendulum, and the linear time-dependent fluctuations 
can be expressed analytically in terms of elliptic functions \cite{PhysRevA.8.2514}.%
\footnote{Similar configurations have been recently shown to be relevant in QCD and 
referred to with the name \emph{Chiral Soliton Lattices} \cite{Brauner:2016pko,Brauner:2019aid}.}

\begin{figure}[h]
\begin{center}
\includegraphics[width=0.49\textwidth]{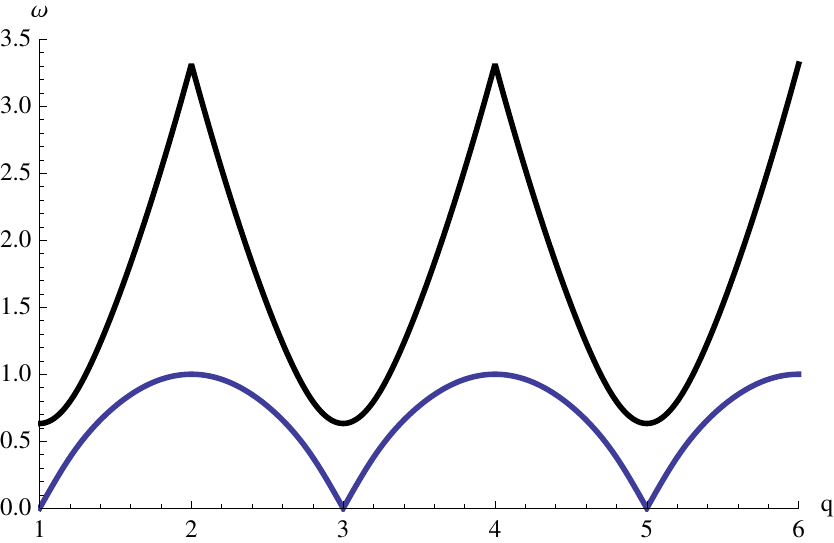}
\end{center}
\caption{Lower portion of the extended Brillouin zone of model \eqref{model1} (the present graph highlight the low-energy part of the dispersions lines already depicted in \ref{pho_dis} ).
There is no gap between the lower and the higher bands.}
\label{no_gap}
\end{figure}

Unlike in the comparison with the one-dimensional chain of Subsection \eqref{cha}, 
there is no obvious relation between the model \eqref{model1} and the kink crystal (KC) of \eqref{Hsg}.
Nevertheless, it is interesting to underline some remarkable similarities of the resulting dispersion relations.
Both \eqref{model1} and the KC feature two kinds of bands. The lower band is in both cases of the bouncing type while the upper is concave,
see the optical branch in Figure \ref{com_cha}.
The two bands in the KC are separated by a gap controlled by the sine Gordon mass $\mu$ in \eqref{Hsg} while in \eqref{model1} there is no gap,
see Figure \ref{no_gap}.

\section{A shift-symmetric model}
\label{mzero}

We modify \eqref{model1} setting to zero the mass term and introducing a term with third derivatives,%
\footnote{The new higher-derivative term appeared to be necessary to obtain backgrounds which feature a propagating shifton mode and pass the stability checks.
Further study is necessary to claim its necessity. A similar role of higher derivative terms emerged in \cite{Musso:2018wbv}.}
\begin{align}\label{model2}
 {\cal L} =
 &\ \frac{1}{2} \dot\phi^2
 -\frac{A}{2} \phi'^2
 +\frac{B}{4} \phi'^4
 +\frac{C}{2} \phi'^2 \phi''^2
 + D \phi''^4
 + E \phi'^2 \phi'''^2\ .
\end{align}
The field $\phi$ appears in the Lagrangian only through its derivatives, so constant field shifts are a symmetry of \eqref{model2}.
We consider again the ansatz \eqref{ansa}, thus obtaining the following equation of motion
\begin{equation}
 \begin{split}
 & 3 k^4 \rho^2 \sin^2(k x) \left\{B+2 k^2 \left[C +2 k^2 (5 E -3 D)\right]\right\}\\
 &\qquad -k^2 \left\{A+k^4 \rho ^2 \left[C+12 k^2(E-D)\right]\right\} =0\ ,
 \end{split}
\end{equation}
which is solved by
\begin{align}\label{Aex}
 A &= k^4 \rho^2 \left[12 k^2 (D -E)-C\right]\ ,\\ \label{Bex}
 B &= -2 k^2 \left[C+2 k^2 (5 E-3 D)\right]\ .
\end{align}

\subsection{Stability under geometric deformations}

To check the stability of solution \eqref{ansa} in model \eqref{model2} with respect to static geometric deformations \eqref{geo_def},
we need to extend the analysis of Subsection \ref{stage} to comprehend one order higher in derivatives,
\begin{equation}\label{ene_thi}
 E[\bar \phi] = \int dx\ {\cal E}\left[\bar \phi(x), \partial \bar\phi(x), \partial^2 \bar\phi(x), \partial^3 \bar\phi(x)\right]\ ;
\end{equation}
the stability check remains nevertheless analogous. Up to boundary terms%
\footnote{We are understanding some IR regularisation provided either by periodic boundary conditions (after a large number of unit cells) or a slow exponential damping.}, one can ``diagonalise'' the quadratic variation ${\cal E}_2$ of \eqref{ene_thi}
\begin{equation}\label{E2}
 {\cal E}_2 = e_{33}\, (\partial^3\xi)^2 + e_{22}\, (\partial^2\xi)^2 + e_{11}\, (\partial\xi)^2\ ,
\end{equation}
where
\begin{align}\label{d33_ext}
 e_{33} &= -E k^4 \rho ^4 \sin ^4(k x)\ ,\\ \label{d22_ext}
 e_{22} &= -\frac{1}{2} k^4 \rho ^4 \sin ^2(k x) \left[\left(C+26 E k^2\right) \sin ^2(k x)+12D k^2 \cos ^2(k x)\right]\ ,\\ \label{d11_ext}
 e_{11} &= \frac{1}{16} k^6 \rho ^4 \left\{-3 \cos (4 k x) \left[C+4 k^2 (5 E-3D)\right] \right.\\
 &\qquad \left.+4 \left(C+24 E k^2\right) \cos (2 k x)-C-12 k^2 (19D+3 E)\right\}\ .
\end{align}
Again, the stability check is passed when the diagonal coefficients are locally positive throughout the entire unit cell.
See Figure \ref{geo_stab} for an explicit example.

\subsection{Fluctuations}

The ansatz \eqref{ansa}, when considered as a solution for model \eqref{model2}, breaks both translations and shift symmetry. 
We thus expect to have a massless mode both around $q=0$ (the \emph{shifton}) and around $q=k$ (the \emph{phonon}).%
\footnote{This is point is further commented in Subsection \ref{tra_shi}.}
The quadratic Fourier Lagrangian is still of the form \eqref{fur} but with the following coefficients: 
\begin{align}\label{coe_fun_shift}
a_0(k,\omega,q) &= \frac{\pi}{2}   k^2 q^2 \rho ^2 \left(q^2 \left(C+4 k^2 (3 D+2 E)\right)-3
   C k^2+2 k^4 (6 D-17 E)+2 E q^4\right)+\pi  \omega ^2 \\ 
a_+(k,\omega,q) &=\frac{\pi}{4}  \left[C-12 D k^2+2 E \left(7 k^2+2 k q+q^2\right)\right]  k^2  \rho ^2 q (k-q) (2 k+q) (3 k+q)\ ,\\
a_-(k,\omega,q) &=-\frac{ \pi}{4} \left[C-12 D k^2+2 E \left(7 k^2-2 k q+q^2\right)\right] k^2 \rho ^2 q (2 k-q) (3 k-q) (k+q) \ .
\end{align}
Since the matrix \eqref{matriciana} associated to the quadratic Lagrangian 
connects only wavevectors which differ by even multiples of $k$,
the modes about $q=0$ and those about $q=k$ can be studied separately and
the shifton and phonon  sectors ``decouple". 

The study of the linear fluctuations is analogous to that of Subsection \ref{flu}.
In particular, the qualitative structure of the matrix $M$ is the same. 
There are though some technical differences, both the new higher coupling $E$ and $m=0$ affect the 
explicit expressions of the entries of $M_{[N]}$.
Looking at $q=k$, there is a $1\times 1$ diagonal block (unlike the $2\times 2$ block found at $q=0$ while studying the phonon):
\begin{equation}
a_0(k,\omega,0) = \pi\omega ^2 \ .
\end{equation}
It shows that $\omega=0$ is a zero of the determinant of the matrix \eqref{matriciana}, hence, it corresponds to a massless mode. 
This corresponds to the Nambu-Goldstone mode coming from the spontaneous symmetry breaking of the shift symmetry. 
The determinant of $M_{[N]}$ around $q=0$ (we take $q=p$ where $p$ is small) can be written as  
\begin{equation}\label{dette2}
 \begin{split}
& ({\cal A}\omega^2 + {\cal B}_{[N]} p^2 + {\cal C}_{[N]} p^4 + ...) \cdot ({\cal H}_{[N]} + {\cal I}_{[N]} \omega^2 + {\cal L}_{[N]} p^2 + {\cal M}_{[N]} p^4 + ...) \cdot ...\ ,
 \end{split}
\end{equation}
where the coefficient ${\cal A}$ does not depend on $N$, while the other coefficients do.
The squared propagation velocity of the shifton is given by
\begin{equation}
 \omega^2 = -\frac{{\cal B}_{[N]}}{{\cal A}} p^2 = c^2_{[N]} p^2 + ...\ ,
\end{equation}
and can be computed in terms of
\begin{equation}\label{speed2}
c^2_{[N]} = -\frac{c^{(0,2)}_{[N]}}{c^{(2,0)}_{[N]}}\ ,
\end{equation}
where $c^{(n,m)}_{[N]}$ is defined to be the coefficient of the term $O(\omega^n,p^m)$ of the determinant \eqref{dette2} of $M_{[N]}$.%

\newpage
\subsection{Explicit example}

We consider the specific case%
\footnote{If we just take $m=0$ in \eqref{cas_mas}, so considering $m=0$, $C=-1$,
$D=-1/10$ and $E=0$, we find a case that (despite passing the stability test against static geometric deformations) features an imaginary propagation speed for the phonon.}.
\begin{figure}[t]
\includegraphics[width=0.5\textwidth]{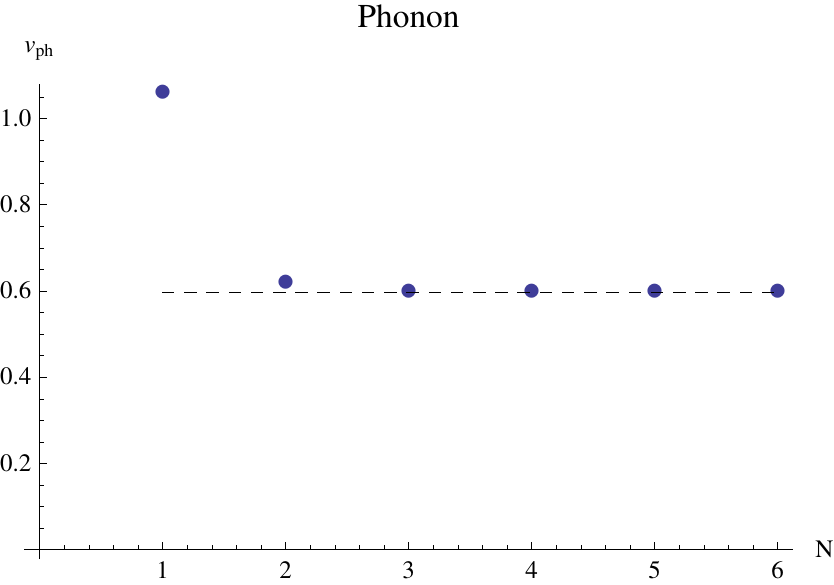}
\includegraphics[width=0.5\textwidth]{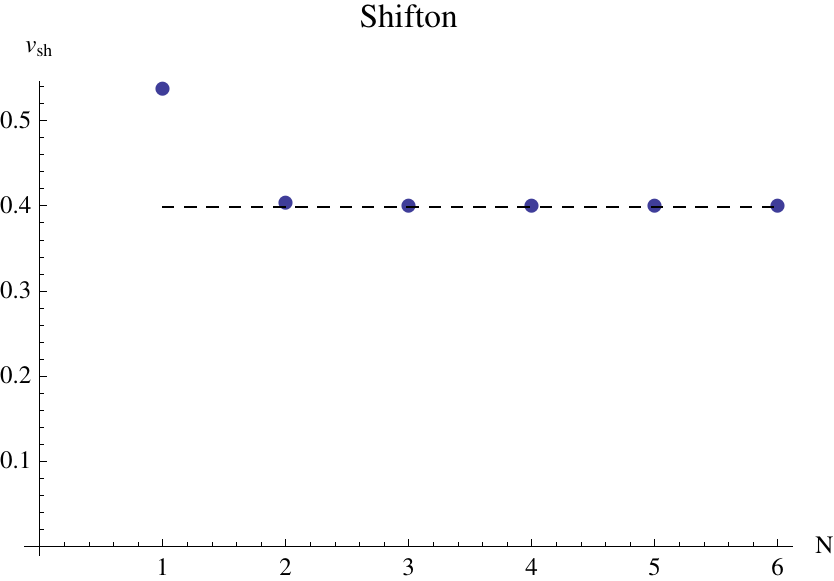}
\caption{Phonon and shifton propagation speeds in the case specified by \eqref{expl_ext1} and \eqref{expl_ext2} as a function of the truncation level $N$
(see discussion below \eqref{matriciana}).
The actual speed values are $v_{\text{ph}}=0.5967...$ and $v_{\text{sh}}=0.3986...$.}
\label{fin}
\end{figure}
\begin{equation}\label{expl_ext1}
C= 0\ , \qquad 
D= -1\ , \qquad 
E= -\frac{1}{10}\ ,
\end{equation}
and determine $A$ and $B$ using \eqref{Aex} and \eqref{Bex} upon requiring to have a solution for
\begin{equation}\label{expl_ext2}
k=1\ , \qquad
\rho = 1\ .
\end{equation}

In order to study the propagation speed of the shifton and phonon modes,
we use the same strategy as described in Section \ref{flu} expanding the determinant of \eqref{matriciana}
around $q=0$ and $q=k$, respectively.
The result for both modes is a propagation speed converging very rapidly to a finite value; we show the values corresponding to $N=1,2,3,4,5,...$
\begin{equation}
\begin{split}
 v_{\text{ph}} &= \left\{\sqrt{\frac{47}{42}},\frac{\sqrt{\frac{85229}{55986}}}{2},\frac{\sqrt{\frac{25986427}{18236773}}}{2},\frac{\sqrt{\frac{588403896531}{413122065094}}}{2},\sqrt{\frac{125394687366827}{352161814236342}}\, ...\right\}\\
 &= \left\{1.05785,\ 0.616913,\ 0.596856,\ 0.596717,\ 0.596717\ ... \right\}\ ,
\end{split}
\end{equation}
and
\begin{equation}
\begin{split}
 v_{\text{sh}} &= \left\{\frac{\sqrt{\frac{201}{7}}}{10},\frac{\sqrt{\frac{32219}{7982}}}{5},\frac{\sqrt{\frac{1811717}{114006}}}{10},\frac{7
   \sqrt{\frac{92017}{126102}}}{15},\frac{\sqrt{\frac{27239575213}{1714117929}}}{10},\ ...\right\}\\
 &= \left\{0.535857,\ 0.401819,\ 0.39864,\ 0.398639,\ 0.398639,\ ... \right\}\ ,
\end{split}
\end{equation}
see also Figure \ref{fin}.
\begin{figure}[t]
\begin{center}
\includegraphics[width=0.49\textwidth]{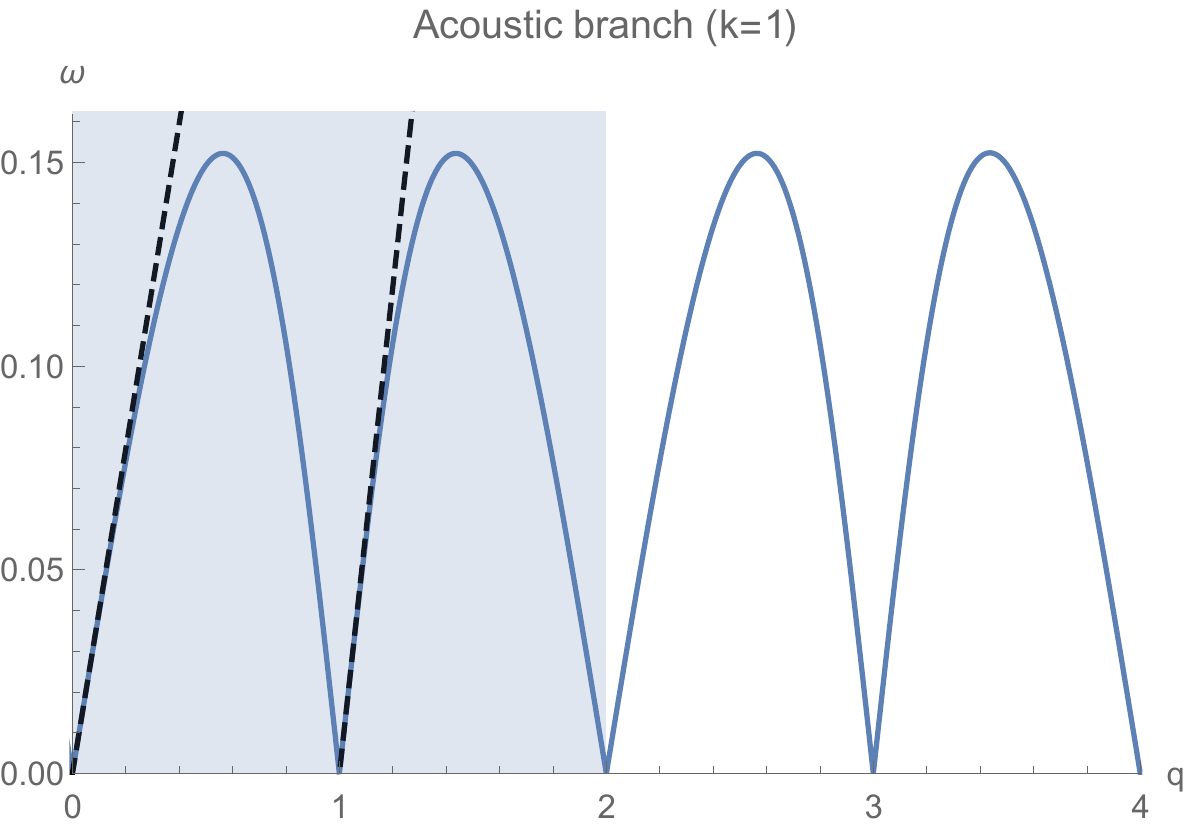}
\includegraphics[width=0.49\textwidth]{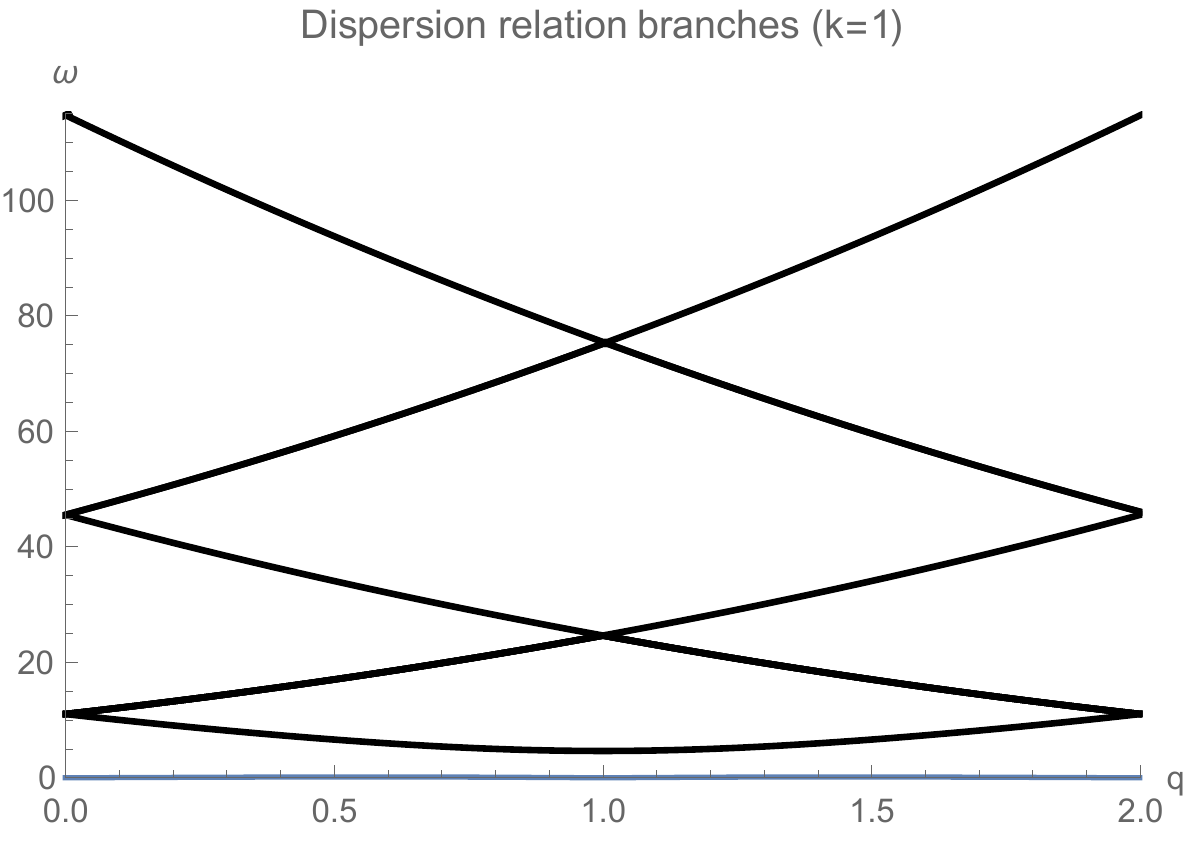}
\end{center}
\caption{Spectral structure of the shift-symmetric model \eqref{model2} in the case \eqref{expl_ext1} and \eqref{expl_ext2}.
Left: acoustic branch featuring shifton ($q\sim 0 + 2 k$) and phononic ($q\sim k + 2k$) linear dispersion regions.
The shiftonic propagation speed is $v_{\text{sh}}=0.399...$ while the phononic speed is $v_{\text{ph}}=0.597...$.
Right: Tower of optical modes; the flat blue line coincides with the bouncing dispersion curve of the left plot.}
\label{bou_shi}
\end{figure}

In Figure \ref{bou_shi} we see that the acoustic phonon branch has developed shiftonic dips.
The overall periodicity is still $2k$, but we have light modes for any integer multiple of $k$.

\section{Result, discussion and comments}
\label{discu}

\subsection{Translation symmetry breaking}

By providing an explicit counterexample, the analysis presented above allows us to drive some generically valid conclusion about translation symmetry breaking:
neither a finite charge density nor shift symmetry are necessary to the purpose of breaking translations dynamically.
More generically, model \eqref{model1} breaks translations and does not possess any extra continuous symmetry.

This point is interesting because it contrasts some generic expectation arising from existing effective approaches.
A generic class of effective field theories for fluids, membranes and elastic media adopts scalar fields that parametrise the coordinates of the target space.
In such a schematisation, the shift symmetry of the scalars corresponds by construction to the translation symmetry of the ambient space and it is thereby unavoidable.
In fact, in such effective models, the breaking of translations corresponds to the locking between the target space translations (\emph{i.e.} the shift symmetry)
and the translations on the manifold on which the scalars live.
Similarly, Q-lattice constructions rely on the presence of a global symmetry under which the low-energy field transform; 
the product of translations times the global symmetry is broken to the diagonal subgroup.

The effective constructions that rely on a global symmetry of the low-energy fields identify 
the global symmetry and translations, this implies that the effective fields are directly the Nambu-Goldstone modes of the symmetry breaking.
In the models of the present paper, instead, the field $\phi$ is not directly the Nambu-Goldstone and it does not parametrise the flat direction of degenerate vacua.
Rather, the dynamics of $\phi$ is responsible for the symmetry breaking itself; its small fluctuations about the non-trivial vacua will then encode (in a non-trivial way) the 
Nambu-Goldstone content as well as other low-energy modes (\emph{e.g.} the optical branches, see Subsection \ref{pc}).

It is important to stress that the present study proves that the cosinusoidal solutions \eqref{ansa} are minima of the static energy, yet they could not be global minima.
Also local minima can give rise to gapless modes interpretable in the framework of (generalised) Goldstone expectation.
Besides, real-world crystals are not always ground states, there are many which are just metastable. 
A famous example is diamond, which is metastable towards decaying to grafite.

\subsection{Absence of extra constraints}
\label{tuning}

Equations \eqref{ro} and \eqref{ka} give the parameters of the ansatz \eqref{ansa}
in terms of the coefficients in the Lagrangian \eqref{model1}.
Apart from the requirement of positivity of the radicands, the couplings $A, B, C, D$ and $m$ are unconstrained. 
An analogous observation holds for the shift-symmetric model \eqref{model2} too.

Generic extensions of models \eqref{model1} and \eqref{model2} would in general require specific relations among the new couplings in the Lagrangian
in order to admit harmonic solutions of the form \eqref{ansa}. 
Nonetheless, by relaxing the ansatz \eqref{ansa}, one can undertake a wider
exploration of models possessing spatially-modulated, anharmonic solutions which could prove to exist without requiring fine-tuning. Likely, such systematic and generic exploration requires numerical approaches.\\
\color{black}

\subsection{Translation and shift symmetry}
\label{tra_shi}

Model \eqref{model2} enjoys shift symmetry $\phi \rightarrow \phi + c$ with $c$ constant. 
Shift symmetry is broken spontaneously by the ansatz \eqref{ansa}, as it would be for \emph{any} field configuration. 
The spontaneously broken shift symmetry corresponds to a zero mode $ \Delta_\alpha \phi(x)$ that is just a constant given by%
\footnote{
Note that the boundary conditions of the shift zero mode are different from those necessary to study the translation zero mode \eqref{tra_zer_mod}.
}
\begin{equation}
 \Delta_\alpha \phi(x) = \alpha\, [P_s,\phi(x)] = \alpha\ ,
\end{equation}
where $P_s$ denotes the generator of the shifts,
\begin{equation}
P_s = \frac{i}{\hbar}\int dx\, \pi(x)\ ,
\end{equation}
and $\pi(x)$ is the canonical conjugated variable to $\phi(x)$ (as in \eqref{coni}).

Consider the generic infinitesimal and local variation of the field $\phi(x)$ under the combination of a shift and a translation (this latter generated by $P_x$), namely
\begin{equation}\label{var}
\delta \phi(x) = \left[\Big(\xi(t,x) P_x + \alpha(t,x) P_s\Big), \phi(x)\right]\ ,
\end{equation}
where the fields $\xi(t,x)$ and $\alpha(t,x)$ are normalisable and correspond to the phonon and the shifton in a coset construction.
The phonon and the shifton are not independent because they can compensate each other; more precisely,  $\delta\phi(x)=0$ has non-trivial solutions
for $\xi(t,x)$ and $\alpha(t,x)$
\begin{equation}\label{inv_hig}
\delta\phi(x) = \xi(t,x) \phi'(x) + \alpha(t,x) = 0\ ,
\end{equation}
where we have considered the explicit action of the generators on the field $\phi(x)$:
\begin{equation}
\left[P_x, \phi(x)\right] = \phi'(x)\ , \qquad 
\left[P_s, \phi(x)\right] = 1\ .
\end{equation}
Condition \eqref{inv_hig} (sometimes called \emph{inverse Higgs constraint}) shows that the parametrisation of the coset by independent phonon and shifton field is redundant \cite{Nicolis:2013sga}.
Shifton and phonon are in this strict sense not physically independent modes.
This is not in tension with our results obtained studying the model \eqref{model2}; 
in fact, the phonon and the shifton arise as a useful effective descriptions of different portions of the same dispersion relation, well separated in momentum space (assuming $q\ll k$ where 
$q$ is the Fourier momentum of the linear fluctuations as in \eqref{fur}). To help intuition, we find ourselves in an analogous situation to phonons and rotons in superfluid helium \cite{Nicolis:2017eqo},
see Subsection \ref{pc} for further comments on this.

\subsection{``Particle" content and first optical branch}
\label{pc}

The field theory models studied above feature a single real scalar field.
The mode content is however rich, because different dispersion branches (or different portions thereof)
admit an effective description in terms of different ``particles". At the lowest energies, the lower bouncing branch has spectral weight in the vicinity 
of the multiples of $2k$ in model \eqref{model1}, corresponding to acoustic phononic modes. In the shift symmetric model \eqref{model2},
the periodicity of the lowest branch is doubled, and at odd multiples of $k$ we find spectral weight contributed by the shiftonic modes.

Raising the energy, and still focusing on the acoustic branch, the simple picture in terms of particles becomes less natural because the lowest branch presents points where the 
group velocity vanishes, this however occurs also in standard models for a classical lattice, like \eqref{spr_bal}.
Relatedly, it is useful to compare the acoustic branch of the shift symmetric model \eqref{model2} 
to the dispersion relation arising in superfluid helium (see for instance \cite{Nicolis:2017eqo});
there the dispersion relation features a maximum too. 
In superfluid helium, while on the left of the maximum there is a low-energy linear portion which corresponds to superfluid phonons,
on the right there is a dip ending into a finite minimum. 
The portion of the dispersion relation close to the relative minimum is interpreted as giving rise to gapped particles called rotons \cite{Landini,2017PhRvL.119z0402C}.
The rotons do not descend from symmetry principles and have the same quantum numbers of the phonons 
(in fact we are just speaking of two regions of the same dispersion relation).

A roton-like dip in the dispersion relation can represent the hint of a nearby modulated phase \cite{Brauner:2019aid}. 
So, inverting the logic of the previous paragraph, the phonon we are studying can be thought as a gapless roton-like dip within a shiftonic dispersion curve. This argument confirms the idea that the phonon arises at a point where $\omega=0$ and $q\neq 0$, where the non-zero value of the wave vector is related to the vacuum periodicity.

The first optical branch can be expected \emph{a priori} and its mass can be computed on the basis of the rigid deformations of the modulated vacuum.
Equation \eqref{tra_zer_mod} defines a rigid zero-mode corresponding to translations of the vacuum as a whole; such zero-mode complies with the periodicity of the vacuum 
and corresponds to choosing periodic Dirichlet boundary conditions. There is an alternative sector that complies with the periodicity of the background but entails Neumann boundary conditions;
this corresponds to a rigid mode of the form $\zeta \cos(k x)$, where $\zeta$ is an infinitesimal parameter. The second variation of the static energy with respect to $\zeta$ gives again 
the mass $m_1^2$ (already obtained in \eqref{m1} in a different way) which characterises the gap of the first optical branch. 
Similarly to the acoustic branch, also the first optical mode can be interpreted as promoting the parameter $\zeta$ to be a normalisable deformation of the corresponding (gapped) rigid mode.

\subsection{UV cut-off}
\label{UVc}

The simple scalar field models considered in the present paper are able to get standard results of spatially periodic systems from a generalised first-principle derivation.
The qualitative agreement with the one-dimensional chain and the presence of phonons upon breaking spatial translations are such examples.
One important difference with respect to standard discretised models is precisely that the field theories studied here are not discretised.
Even if the solutions feature a modulation characterised by a wavevector $k$, there is no UV cut-off associated to such a physical scale, at least at the classical level.

Another important difference is that the equation of motion for the fluctuations is not analogous to Schr\"odinger equations for lattice potentials,
indeed it is not given by a second-order derivative term plus a periodic potential; rather we have a non-canonical, higher-order derivative term and a space-independent potential.
As a consequence Bloch's (or Floquet's) theorem does not apply trivially to the present case (see for instance \cite{flo}).

An important future direction consists in quantising the model. The evaluation of the actual validity range in energy of the enriched effective description we proposed
and the possibility of turning on a finite temperature avoiding UV catastrophes are two main questions which remain open.
As an intermediary step, it would also be necessary to study the interactions and thus to go beyond the linear approximation 
used in the present paper to characterise the spectrum (in the language of crystal dynamics this is related to anharmonicities).

\subsection{Future perspectives}

In addition to the perspectives mentioned earlier in this section, there are some further particularly interesting future directions.

The models of the present paper undertake the quest for the simplest field-theoretical framework in which it is possible to break translations.
Such minimality could be a promising factor in view of a wider applicability of the results. 
However, model \eqref{model1} descends from specific choices; it would therefore be relevant to assess the potential genericity 
of the gradient Mexican hat mechanism by means of a systematic analysis of the most generic effective model given in Appendix \ref{gen_ter}.

The gradient Mexican hat construction is clearly based on the presence of non-linearities in the field and in the spatial derivatives thereof.
Non-linearities have been recently considered in the context of photonic lattices \cite{2017arXiv170702213A,PhysRevLett.116.163902} and have been proven
to crucially lead to either uniform or modulated phase patterns depending on their defocusing/focusing character.
Specifically, \cite{2017arXiv170702213A} shows that the sign of a particular non-linear term (the so-called Kerr term) in the Schr\"odinger 
equation for monochromatic light destabilises the uniform vacuum and favours chessboard phase configurations.
There is thus an analogy with the Mexican hat constructions where the appropriate signs of the various terms in the Lagrangian make them compete;
it is in fact the optimal compromise of such competition which defines the non-trivial vacuum.
It would be interesting to make the parallel in a more precise way, and explore the possibility of implementing Mexican hat gradient terms in actual experiments.

All the models studied in the present paper are in $(1+1)$-dimensions, nonetheless they allow for higher dimensional generalisations.
Apart from exploring the working of higher dimensional gradient Mexican hats and the stability of the corresponding vacua,
the increase of the dimensionality of the system has an obvious interest in order to circumvent the no-go theorems for symmetry breaking in low-dimensionality
\cite{Mermin:1966fe,Coleman:1973ci,Gross:1974jv}.
It is possible that the present models serve as an effective description for some holographic systems,
in this case the destabilising fluctuations which in low-dimensions prevent the ordering are suppressed by large-$N$ effects \cite{Witten:1978qu,Anninos:2010sq,Argurio:2016xih,Argurio:2019qcc}.

The model specified in \eqref{model1} only breaks translations, hence it is not a \emph{supersolid} \cite{2012RvMP}.
 However, inhomogeneous translation breaking in a model with a complex scalar field subject to an extra $U(1)$ symmetry could possibly describe a supersolid.
 It also interesting to ask whether the shift symmetry of \eqref{model2} could have a role in this respect.

Eventually, models for scalar field dark energy (see for instance \cite{ArkaniHamed:2003uy,Afshordi:2006ad}) play with non-trivial modulated vacua 
of simple scalar field theories. It would be interesting to explore the connections to the present study, especially in relation to their thermodynamic properties.

\section{Acknowledgments}

It is a pleasure to acknowledge Riccardo Argurio for guidance at an early stage of the project.
We thank Daniel Are\'an, Anxo Biasi, Carlos Hoyos, Javier Mas, Giorgio Musso, V\'ictor Pardo, \'Angel Paredes and Alfonso Ramallo for relevant discussions and observations.
\vspace{10pt}

The work of DM has been funded by the Spanish grants FPA2014-52218-P and
FPA2017-84436-P by Xunta de Galicia (GRC2013-024), by FEDER 
and by the Mar\'ia de Maeztu Unit of Excellence MDM-2016-0692

\appendix 

\section{Energy-momentum tensor, general case}
\label{stress}

The generic equation of motion for a higher-derivative Lagrangian is
\begin{equation}\label{eom_gen}
\sum_{i=0}^n (-1)^i \partial_{\alpha_1}...\partial_{\alpha_i} \frac{\delta {\cal L}}{\delta \partial_{\alpha_i}...\partial_{\alpha_n} \phi}
= 0\ ,
\end{equation}
where $n$ is the maximum number of derivatives applied on the single field in the Lagrangian.

Consider the infinitesimal diffeomorphism generated by $\delta_\xi$ where $\xi^\mu(x)$,
its action on the scalar field $\phi$ is given by
\begin{equation}
 \delta_\xi \phi = -\xi^\mu \partial_\mu \phi\ .
\end{equation}
The Lie variation of the Lagrangian is given by
\begin{equation}\label{symme}
\begin{split}
\delta_\xi^{(\text{Lie})} {\cal L} =& \
\delta_\xi {\cal L}
-\sum_{i=0}^n \frac{\delta{\cal L}}{\delta \partial_{\alpha_1}...\partial_{\alpha_i}\phi}\delta_\xi \partial_{\alpha_1}...\partial_{\alpha_i}\phi\\
=& \
-\xi^\mu \partial_\mu {\cal L}
-\sum_{i=0}^n  \frac{\delta {\cal L}}{\delta \partial_{\alpha_1}...\partial_{\alpha_i} \phi} \delta_\xi \partial_{\alpha_1}...\partial_{\alpha_i} \phi\\ 
=& \
-\xi^\mu \partial_\mu {\cal L}
-\sum_{i=0}^n  \frac{\delta {\cal L}}{\delta \partial_{\alpha_1}...\partial_{\alpha_i} \phi}  \partial_{\alpha_1}...\partial_{\alpha_i} \delta_\xi \phi
  = 0\ ,
  \end{split}
\end{equation}
where we have required it to be zero to enforce translation symmetry.
By using the equation of motion \eqref{eom_gen} and setting $\xi^\mu(x)$ to be a constant at last,
Equation \eqref{symme} can be written as
\begin{equation}
 \partial_\mu T^\mu_{\ \nu} = 0\ ,
\end{equation}
where%
\footnote{Expression \eqref{em} is computed also in \cite{Ilin:2018anz}.}
\begin{equation}\label{em}
\begin{split}
T^\mu_{\ \nu} =&\ \delta^\mu_\nu {\cal L}
- \sum_{i=0}^{n-1} \sum_{j=0}^i (-1)^{j}\, \partial_{\alpha_1}...\partial_{\alpha_j} \frac{\delta {\cal L}}{\delta \partial_{\mu}\partial_{\alpha_1}...\partial_{\alpha_i}\phi} \partial_{\alpha_{j+1}}...\partial_{\alpha_i} \partial_\nu \phi
\ .
\end{split}
\end{equation}
Despite their covariant aspect, the expressions above do not assume the relativistic invariance of the system.

\section{Recursion structure}
\label{rec_stru}

We consider the generic $N\times N$ block-tridiagonal matrix
\begin{equation}
 A_N = \left( \begin{array}{cccccc}
 a_{1,1} & a_{1,2} & 0 & 0 & 0& ...\\
 a_{2,1} & a_{2,2} & a_{2,3} & 0 & 0& ...\\
 0& a_{3,2} & a_{3,3} & a_{3,4} & 0&... \\
 ...& ... &... &...& ...&...
 \end{array}\right)\ .
\end{equation}
The determinant of $A_N$ can be obtained recursively in terms of the determinants of $A_M$ with $M<N$, specifically
\begin{equation}\label{det_rec}
 f_N \equiv \det\left( A_N \right) = f_{N-1} a_{N,N} - a_{N,N-1}a_{N-1,N} f_{N-2}\ .
\end{equation}

Following \cite{NLA}, it is convenient to express $A_N$ as
\begin{equation}
 A_N = D_{A_N} - L_{A_N} - U_{A_N}\ ,
\end{equation}
where
\begin{equation}
 D_{A_N} = \left( \begin{array}{cccccc}
 a_{1,1} & 0 & 0 & 0 & 0& ...\\
 0 & a_{2,2} & 0 & 0 & 0& ...\\
 0& 0 & a_{3,3} & 0 & 0&... \\
 ...& ... &... &...& ...&...
 \end{array}\right)\ ,
\end{equation}
\begin{equation}
 U_{A_N} = -\left( \begin{array}{cccccc}
 0 & a_{1,2} & 0 & 0 & 0& ...\\
 0 & 0 & a_{2,3} & 0 & 0& ...\\
 0& 0 & 0 & a_{3,4} & 0&... \\
 ...& ... &... &...& ...&...
 \end{array}\right)\ ,
\end{equation}
and
\begin{equation}
 L_{A_N} = -\left( \begin{array}{cccccc}
 0 & 0 & 0 & 0 & 0& ...\\
 a_{2,1} & 0 & 0 & 0 & 0& ...\\
 0& a_{3,2} & 0 & 0 & 0&... \\
 ...& ... &... &...& ...&...
 \end{array}\right)\ .
\end{equation}
Defining
\begin{align}
 L_N \equiv D_N - L_{A_N}\ ,\\
 U_N \equiv D_N - U_{A_N}\ ,
\end{align}
the matrix $A_N$ is diagonalised as follows
\begin{equation}
 D_N^{-1} = L_N^{-1} A_N U_N^{-1}\ ,
\label{expressionInvD}
\end{equation}
where entries of the diagonal matrix $D_N$ are given by
\begin{equation}
 (D_N)_{ij} = \delta_{ij} g_i\ ,
\end{equation}
and the function $g_i$ is recursively defined
\begin{align}\label{recurso}
 g_1&=a_{1,1}\ ,\\
 g_n& =a_{n,n} - \frac{a_{n,n-1}a_{n-1,n}}{g_{n-1}}\ .
\end{align}

Thanks to \eqref{expressionInvD} and the definition of $L_N$ and $U_N$, we have that $\text{det}(A_N)=\text{det}(D_N)$, if $D_N$ is invertible. 
We can use $\text{det}(D_N)$ to probe the dispersion relations of the model by assuming a continuous behaviour 
of $\text{det}(A_N)$ with respect to $\omega$.%
\footnote{$\lim\limits_{\omega \to \omega(q)}\text{det}(D_N)\equiv \lim\limits_{\omega\rightarrow \omega(q)}\text{det}(A_N)=\left.\text{det}(A_N)\right\vert_{\omega = \omega(q)}$}

The diagonal entries $g_i$ do not provide individually the dispersion branches of the model.
These latter correspond to the zero's of the entire determinant of $M_{[N]}$ \eqref{matriciana}, $\text{det}M_{[N]} = \prod_i g_i$, 
where cancellations among the zeros and the poles of the individual $g_i$ may occur.
In fact, comparing to \eqref{det_rec} we have
\begin{equation}
 g_n = \frac{f_n}{f_{n-1}} = \frac{\det(A_n)}{\det(A_{n-1})}\ .
\end{equation}

\section{Propagation speed, alternative method}
\label{alte}
Recalling the definition of the coefficient functions $a_{\pm}$ and $a_0$ \eqref{coe_fun}, the recursive function for the eigenvalues \eqref{recurso} takes the form
\begin{align}
 g_1 &= a_0\left[2Nk+q\right]\\
 g_n &= a_0\left[2(N+1-n)k+q\right] \\
 &\qquad - \frac{a_-\left[2(N+2-n)k+q\right]a_+\left[2(N+1-n)k+q\right]}{g_{n-1}}\ .
\end{align}
The determinant of $M_{[N]}$ \eqref{matriciana} is given by $\text{det}M_{[N]} = \prod_i g_i$.
It is possible to get (very accurate) approximate expressions considering just a few terms of the product $\prod_i g_i$
around the entry of $M_{[N]}$ corresponding to the mode studied. 

\section{General model up to $\phi^4$ and $\partial_x^8$}
\label{gen_ter}

We consider the most general model having standard kinetic term $\frac{1}{2}\dot\phi^2$ which comply with the parity requirements $\phi \leftrightarrow -\phi$ and $\partial_x \leftrightarrow -\partial_x$.
Modulo total derivatives, the most general Lagrangian is obtained composing the following terms:
\begin{equation}
\hspace{-8pt}
 \begin{array}{c|ccccccc}
 \text{order in $\partial_x$}&&&&&&&\\ 
 0&\phi^2&{\color{gray}\phi^4}&&&&&\\
 2&\phi'^2&{\color{gray}\phi'^2\phi^2}&&&&&\\
 4&\phi''^2&\phi'^4&{\color{gray}\phi''^2\phi^2}&&&&\\
 6&{\color{gray}\phi'''^2}&{\color{gray}\phi'''^2\phi^2}&{\color{gray}\phi''^3\phi}&\phi''^2\phi'^2&&&\\
 8&{\color{gray}\phi''''^2}&{\color{gray}\phi''''^2\phi^2}&{\color{gray}\phi'''^2\phi'^2}&{\color{gray}\phi'''^2\phi''\phi}&{\color{gray}\phi'''\phi''^2\phi'}&\phi''^4 &
 \end{array}
\end{equation}
The terms written in black are those considered in model \eqref{model1}.

\section{Pressure}

The microscopic definition of the pressure is given by $T_{xx}$ which we have computed explicitly \eqref{Txx}.
The requirement of having a consistent thermodynamic derivation of the pressure
constrains the behaviour of the model under changes of the volume.
In particular, the scalings dictated by naive dimensional analysis do not lead to a thermodynamic picture consistent with the microscopic picture.

Assume that the various quantities rescale according to 
\begin{equation}\label{ska}
X \to a^{r_{(X)}}X\ ,
\end{equation}
under a rigid dilatation by $a=1+\eta$, where $\eta$ is an infinitesimal parameter;
$r_{(A)}$ and $r_{(B)}$ are fixed by the equations of motion \eqref{eom_con_1} and \eqref{eom_con_2} in terms of $r_{(k)}$ and $r_{(\rho)}$.

The internal energy $E$ can be expressed as the energy density times the volume, $E=\epsilon V$,
where the energy density $\epsilon=\bar T_{tt}$ is computed by considering the spatial average of $T_{tt}$ in \eqref{ene_den}.
The generic behaviour of the energy under an infinitesimal rescaling by $\eta$ is controlled by the derivative 
\begin{equation}\label{ene_res}
\frac{dE}{d\eta} = \frac{d\epsilon}{d\eta}V + \epsilon\frac{dV}{d\eta}\ ,
\end{equation}
whose explicit expression depends on the scaling rules \eqref{ska}.

By comparing the behaviour under infinitesimal rescalings, we ask ourselves under which conditions the energy can be expressed by the standard homogeneous form $E=-pV$,
valid at $T=\mu=0$.
To this purpose, it results that the scaling rules \eqref{ska} must satisfy the following system of equations:
\begin{equation}\label{req_sys}
r_{(k)} = -1\ , \qquad 
r_{(C)} = 6-4r_{(\rho)}\ , \qquad 
r_{(D)} = \frac{32}{3} - 4 r_{(\rho)}\ .
\end{equation}
It is remarkable to note that naive dimensional analysis does not satisfy \eqref{req_sys}.%
\footnote{
Considering the spacetime dimensional analysis with $[t]=[x]=1$ we have 
\begin{equation}\label{spacetime} 
 r_{(k)} = -1\ , \qquad 
 r_{(m)} = -1\ , \qquad 
 r_{(\rho)} = 0 \ , \qquad 
 r_{(C)} = 4 \ , \qquad 
 r_{(D)} = 6\ .
\end{equation}
Considering instead spatial dimensional analysis $[t]=0$ and $[x]=1$ we get
\begin{equation}\label{spatial}
 r_{(k)} = 0\ , \qquad 
 r_{(m)} = 0\ , \qquad 
 r_{(\rho)} = -\frac{1}{2} \ , \qquad 
 r_{(C)} = 1 \ , \qquad 
 r_{(D)} = 1\ .
\end{equation}
Both \eqref{spacetime} and \eqref{spatial} lead to the same rescaling of \eqref{ene_res} but are incompatible with the requirements \eqref{req_sys}.
Analogous conclusions are reached also for the shift-symmetric model \eqref{model2}.
}
The possibility of having a simple thermodynamic treatment hints at the need to consider ``anomalous dimensions" already on the basis of classical considerations.
An actual quantum treatment is nevertheless necessary to assess the value of the information contained in \eqref{req_sys}.

\end{document}